\documentclass[11pt,conference]{IEEEtran}
\usepackage{cite}
\usepackage{amsmath,amssymb,amsfonts}
\usepackage{algorithm}
\usepackage{booktabs}
\usepackage{algpseudocode}
\usepackage{graphicx}
\usepackage{textcomp}
\usepackage{siunitx}
\usepackage{xcolor}

\def\BibTeX{{\rm B\kern-.05em{\sc i\kern-.025em b}\kern-.08em
    T\kern-.1667em\lower.7ex\hbox{E}\kern-.125emX}}
\begin{document}

\pagestyle{empty}
\onecolumn
\noindent {\large \copyright ~2022 IEEE.  Personal use of this material is permitted.  Permission from IEEE must be obtained for all other uses, in any current or future media, including reprinting/republishing this material for advertising or promotional purposes, creating new collective works, for resale or redistribution to servers or lists, or reuse of any copyrighted component of this work in other works. \\~\\

\noindent This is a peer-reviewed and accepted version of the following in press document.} \\~\\

\noindent {\large S. Davidson, N. McCallan, K. Y. Ng, P. Biglarbeigi, D. Finlay, B. L. Lan, and J. McLaughlin. ``Epileptic Seizure Classification Using Combined Labels and a Genetic Algorithm,'' {\it 21st IEEE Mediterranean Electrotechnical Conference (MELECON 2022) (In Press)}, 2022.}

\twocolumn
\newpage

\title{Epileptic Seizure Classification Using Combined Labels and a Genetic Algorithm}

\author{\IEEEauthorblockN{
		Scot Davidson\IEEEauthorrefmark{1}\IEEEauthorrefmark{3},
		Niamh McCallan\IEEEauthorrefmark{1},
		Kok Yew Ng\IEEEauthorrefmark{1}\IEEEauthorrefmark{2},
		Pardis Biglarbeigi\IEEEauthorrefmark{1},
		Dewar Finlay\IEEEauthorrefmark{1}, \\
		Boon Leong Lan\IEEEauthorrefmark{2}, and
		James McLaughlin\IEEEauthorrefmark{1}}
	\IEEEauthorblockA{
		\IEEEauthorrefmark{1}NIBEC, Ulster University, Jordanstown Campus, Shore Road, Newtownabbey BT37 0QB, UK.}
	\IEEEauthorblockA{\IEEEauthorrefmark{2}Department of Electrical and Computer Systems Engineering and \\Advanced Engineering Platform, School of Engineering, Monash University, Bandar Sunway, Malaysia.}
	\IEEEauthorblockA{\IEEEauthorrefmark{3}Corresponding author. Email: davidson-s18@ulster.ac.uk}
}

\maketitle

\begin{abstract}
	Epilepsy affects 50 million people worldwide and is one of the most common serious neurological disorders. Seizure detection and classification is a valuable tool for diagnosing and maintaining the condition. An automated classification algorithm will allow for accurate diagnosis. Utilising the Temple University Hospital (TUH) Seizure Corpus, six seizure types are compared; absence, complex partial, myoclonic, simple partial, tonic and tonic-clonic models.  This study proposes a method that utilises unique features with a novel parallel classifier --- Parallel Genetic Naive Bayes (NB) Seizure Classifier (PGNBSC). The PGNBSC algorithm searches through the features and by reclassifying the data each time, the algorithm will create a matrix for optimum search criteria. Ictal states from the EEGs are segmented into 1.8 s windows, where the epochs are then further decomposed into 13 different features from the first intrinsic mode function (IMF). The features are compared using an original NB classifier in the first model. This is improved upon in a second model by using a genetic algorithm (Binary Grey Wolf Optimisation, Option 1) with a NB classifier. The third model uses a combination of the simple partial and complex partial seizures to provide the highest classification accuracy for each of the six seizures amongst the three models (20\%, 53\%, and 85\% for first, second, and third model, respectively).

\end{abstract}

\begin{IEEEkeywords}
	Electroencephalography (EEG), Epileptic Seizure, Classification, Naive Bayes, Genetic Algorithm
\end{IEEEkeywords}

\section{Introduction}
Epilepsy is one of the most common neurological disorders in the world \cite{brodie2012fast}, affecting about 50 million people worldwide \cite{AmountOfPplWhoHaveEpilepsy}. Epileptic seizures occur when millions of neurons are synchronously excited, resulting in a wave of electrical activity in the cerebral cortex \cite{tzallas2009epileptic}. Electroencephalography (EEG) is a noninvasive tool that measures cortical activity with millisecond temporal resolution. EEGs record the electrical potentials generated by the cerebral cortex nerve cells \cite{subasi2005classification}. As a result, this tool is commonly used for the analysis and detection of 
seizures \cite{adeli2003analysis}. Epilepsy causes many difficulties in relation to the quality of life for the patient. The International League Against Epilepsy (ILAE) have outlined a list of identified seizures \cite{ILAERevisedGuide}. There are two responses on the patient, categorised as \textit{motor} responses, where the seizures cause involuntary spasms, or \textit{non-motor} responses, which affect the consciousness of the patient hence reducing their ability to pay attention.


Kukker et al. \cite{intro_comp_3} used a fuzzy Q-learning genetic classifier to improve their original features. The signal is converted using empirical mode decomposition (EMD) where the intrinsic mode functions (IMFs) are converted using the Hilbert-Huang Transform (HHT). Nineteen features are extracted from the conversion with a distinct set of annotations created for each of the IMFs generated. The CHB-MIT dataset was used and a classification accuracy of \SI{96.79}{\percent} was achieved.
Ammar et al. \cite{intro_comp_1} used a particle swarm genetic algorithm for a patient specific seizure classification system. This classification model used seizure and nonseizure as the annotations. The features extracted where kurtosis, skewness, and standard deviation. A smaller dataset (Bonn) was used with a support vector machine (SVM) classifier and an accuracy of \SI{98.89}{\percent} was achieved.
These state of the art designs focused primarily on the detection of seizure and nonseizure within the CHB-MIT and Bonn datasets, respectively. However, the classification of different seizure types was not investigated due to insufficient labelling as well as having limited patient-specific data found in these datasets.

This state-of-the-art work proposes a seizure classification system using multiclass parallel classifiers --- Parallel Genetic Naive Bayes (NB) Seizure Classifier (PGNBSC) --- to classify focal onset seizures against other seizure types. As simple partial and complex partial seizures share morphology and exhibit similar responses on the patient, where the main factor being whether the patient remains conscious during the event, the proposed algorithm can discriminate these two seizure types against other seizures by combining them into a focal seizure. In addition, this algorithm is also tested against the Temple University Hospital (TUH) Seizure Corpus due to its extensive labelling of multiple seizure types as well as more patient-specific data compared to the CHB-MIT and Bonn datasets.

This manuscript is organised as follows: Section \ref{Data} explains the dataset used in this research; Section \ref{Feature} describes the proposed PGNBSC classifier algorithm; Section \ref{Results} presents and discusses the results obtained; and Section \ref{Conclusion} provides the conclusions.

\section{Dataset}\label{Data}
The dataset from the TUH Seizure Corpus v1.5.2 (27 May 2020) was used in this study. This dataset is a subset of a much larger EEG corpus \cite{shah2018temple,7002953}. This dataset is divided by the TUH into two sets, defined as training and testing with demographics being made equal across both sets.
The dataset was annotated by a team established by the university based on the signals and the neurologists' reports. It is annotated by giving the seizure start and end time in a text file. This is converted to a sample number and each window of the seizure is extracted between the start and the end samples. The configurations for the training and testing are shown in Table \ref{tab:Config}.
\begin{table*}[t!]
	\centering
	\caption{The breakdown of each label into its training and testing files followed by the amount of windows taken from the original files.}
	\begin{tabular}{crrrrrrrr}
		\toprule
		\textbf{Seizure Type} & \multicolumn{1}{c}{\textbf{Training}} & \multicolumn{1}{c}{\textbf{Testing}} & \multicolumn{1}{c}{\textbf{wTrain}} & \multicolumn{1}{c}{\textbf{wTest}} & \multicolumn{1}{c}{\textbf{Duration\_Tr (s)}} & \multicolumn{1}{c}{\textbf{Duration\_Tt (s)}} & \multicolumn{1}{c}{\textbf{wDuration\_Tr (s)}} & \multicolumn{1}{c}{\textbf{wDuration\_Tt (s)}} \\
		\midrule
		Absence               & 13                                    & 7                                    & 253                                 & 173                                & 495                                         & 357                                         & 455                                          & 311                                          \\
		Complex partial       & 134                                   & 35                                   & 15,373                              & 4629                               & 27,922                                      & 840                                         & 27,671                                       & 1512                                         \\
		Myoclonic             & 2                                     & 1                                    & 717                                 & 10                                 & 1293                                        & 19                                          & 1290                                         & 18                                           \\
		Simple Partial        & 5                                     & 3                                    & 1055                                & 112                                & 1941                                        & 204                                         & 1899                                         & 202                                          \\
		Tonic                 & 19                                    & 9                                    & 243                                 & 396                                & 455                                         & 749                                         & 437                                          & 819                                          \\
		Tonic-Clonic          & 17                                    & 11                                   & 1080                                & 1981                               & 1967                                        & 3582                                        & 1944                                         & 3566                                         \\
		\midrule
		\textbf{TOTAL}        & \textbf{190}                          & \textbf{66}                          & \textbf{18,721}                     & \textbf{7301}                      & \textbf{34,073}                             & \textbf{5751}                               & \textbf{33696}                               & \textbf{6428}                                \\
		\bottomrule \vspace{-2mm} \\
		\multicolumn{9}{l}{\footnotesize{*wTrain: The amount of labels in training that are equal to the corresponding seizure.}} \\
		\multicolumn{9}{l}{\footnotesize{*wTest: The amount of labels in testing that are equal to the corresponding seizure.}} \\
		\multicolumn{9}{l}{\footnotesize{*wDuration: The windowed duration of training (Tr) and testing (Tt) sets.}}
	\end{tabular}
	\label{tab:Config}
\end{table*}
The TUH includes only two pieces of information on the individual patients, which are the age and the gender of each patient. These demographics are negated because age is only a contributing factor in the absence seizure, which already has a unique waveform \cite{ILAERevisedGuide}.

\section{Feature Extraction and Selection}\label{Feature}
\subsection{Preprocessing of Raw EEGs}\label{Pre-process}
To maintain consistency in the dataset, only the 19 channels common to all seizures defined by the International 10--20 System were used \cite{OOSTENVELD2001713}. By doing so, the unwanted channels such as the electrocardiogram (ECG), electromyography (EMG), as well as photo stimulus channels were omitted. The signals were resampled to \SI{250}{\hertz} using an antialiasing low-pass finite impulse response filter as this is the lowest sampling frequency common to all of the seizures.
The \SI{60}{\hertz} line noise was removed with a bandstop infinite impulse response (IIR) filter. The EMD was then performed and the first IMF was taken for the analysis of the signals \cite{intro_comp_3}. The signals were then divided into \SI{1.8}{\second} windows, which was chosen because it reflects on the shortest ictal window of all the different seizure types. 
If a seizure contains only one window, then the window is taken from the centre of the seizure. Each feature was calculated for each channel of the windowed signals.

\subsection{Time Domain Features}
\subsubsection{Standard Deviation}
Ictal episodes have a higher energy compared to the non-ictal episodes. However, some seizures have a lower energy response such as absence, complex partial, and simple partial \cite{weir1965morphology,devinsky1988clinical,gilliam2000adult}. The standard deviation $\sigma$ measures how much energy is created away from the mean, where a higher energy level would naturally produce a higher $\sigma$.

\subsubsection{Shannon Entropy}
The Shannon entropy method is used to measure the chaotic nature of EEGs.

\subsubsection{Kurtosis}
Kurtosis $\gamma_2$ is the degree of ``peakedness'' of a real valued random variable, which is used to determine if $\sigma$ is created from small constant deviations or if large inconsistent deviations are present in the signal \cite{quitadamo2018kurtosis}.

\subsubsection{Hjorth Criteria}
Hjorth parameters measures a signal via three criteria: activity, mobility, and complexity. Activity uses variance as its basis and is therefore not used as a feature. However, mobility $h_{m}$ and complexity $h_{c}$ can be used to measure the mean frequency of a signal and deviation from a pure sine wave, respectively \cite{HOLLER2019106485}.

\subsubsection{Skewness}
Skewness measures the symmetry of the time series data.
\subsubsection{Energy and Nonlinear Energy}
The total energy of the time series data is used because different seizures can have a muscle artefact which typically has a higher energy content. The mean nonlinear energy measures more than the total energy by measuring the fluctuations in energy from the time series data.
\subsection{Fractal Analysis}
Fractals are used to measure the self-similarity of the given time series values. Absence, tonic, and tonic-clonic seizures have rhythms generating self similar waveforms \cite{gilliam2000adult,kobayashi2009spectral}.

\subsubsection{Higuchi Fractal Dimension}
The time series data have to be firstly decomposed into a set of subseries \cite{Khoa2012}.
A curve is then created to measure the similarity of the subseries. The length of the curve can be computed using
\begin{equation}
	L_m \hspace{-1mm}=\hspace{-1mm} \frac{1}{k}\hspace{-1mm}\left[\hspace{-1mm}\left(\sum^{r}_{i=1}\left|x(m\hspace{-1mm}+\hspace{-1mm}i k)\hspace{-1mm}-\hspace{-1mm}x(m\hspace{-1mm}+\hspace{-1mm}(i\hspace{-1mm}-\hspace{-1mm}1) k)\right| \hspace{-1mm}\right)\hspace{-1mm}\frac{N\hspace{-1mm}-\hspace{-1mm}1}{rk}\right],
	\label{Fractal_line_equation}
\end{equation}
where $r = \left\lceil\frac{N-m}{k}\right\rfloor$ and $x(i)$ is the subseries.

\subsubsection{Katz Fractal Dimension}
Katz utilises the same subseries defined above. It is another fractal estimation algorithm where the successive points are measured to compute the self-similarity such that the fractal dimension of the subseries is used as another feature.

\subsubsection{Spectral Entropy}
To calculate the spectral entropy of a time series data, it is initially converted to the frequency domain using fast Fourier transform, which is then further converted into power.
The mean, maximum, and minimum are then taken as the features to be fed into the PGNBSC.
To balance the dataset, an upsampling method is used where the least represented labels are repeated by the nearest integer factor of the largest label. In the first model, the largest label is the complex partial seizure. In the second model, the largest is the complex partial seizure. In the third model, the largest label is the focal-onset.

\subsection{Genetic Algorithms}
Genetic algorithms are optimisation methods that mimic Darwin theory. They are exploratory procedures that seek to find the near optimal solutions to complex problems \cite{genetic_algo}. Emary et al. \cite{emary2016binary} adapted a previous genetic algorithm called Binary Grey Wolf Optimisation (BGWO) to be applied for feature selection. The proposed work in this research uses the BGWO Algorithm 1 from Emary et al. \cite{emary2016binary} because it provides the shortest compilation time and is shown in Algorithm 1.
\begin{algorithm}[b!]
	\caption{Binary Grey Wolf Optimisation}\label{alg:cap}
	$I$ iterations \\
	$X_{i}$ Population Initalisation \\
	$W_1$ = first search agent \\
	$W_2$ = second search agent \\
	$W_3$ = third search agent \\
	$t_i$ = $0$
	\begin{algorithmic}
		\For{$i=1:3$}
		\State $W_i \gets \text{New Global position}$
		\EndFor
		\While{$t_i<I$}
		\State $\text{Apply fitness function to $W_1, W_{2}, W_{3}$}$
		\State $\text{Evaluate $W_1, W_2, W_3$}$
		\State $\text{Compare $W_1, W_2, W_3$}$
		\State $t_i=t_i+1$
		\EndWhile
	\end{algorithmic}
\end{algorithm}

To use a genetic algorithm, the key feature is the fitness function. The fitness function is initialised using a randomiser. This selects genes for the starting lineage of the subsequent pairs, which are defined as $W_{i}$ in Algorithm \ref{alg:cap}. As a fail-safe, should the algorithms selects no features, then the cost is set to $\infty$. The same classifier defined in Section \ref{S:classification} is used to evaluate the features and the F1-Score is used for the comparison performance of the selected features. As parallel classifiers are used, the F1-score can be used without having to use the global F1-score formula. The F1-score is perfect if the score is $1$ however genetic algorithms are perfect if they score zero. Therefore, we use $1-F1$ to manipulate the F1-score towards $0$. To ensure the gene-line and the subsequent gene-lines are improving, a bias is created using the the iteration number. As the algorithm is improving the bias becomes higher meaning the gene-lines must be improving or the new gene-line is entered into the gene-pool. 

An early stopping criterion is also set in the algorithm that measures the six previous values. This is measured by comparing the deviation in the entire signal with the deviation in the final six values. If the deviation is $<5\%$ of the total deviation, then the stopping criterion is finished and the features are extracted.

\subsection{Classification} \label{S:classification}
The classifier chosen is the Naive Bayes classifier. It is a probabilistic classifier based on the Bayes theorem under the assumption that any feature of a particular class is independent of any other feature. Error estimation is computed based on the maximum likelihood \cite{fielding2006cluster}. The prior probabilities are set to uniform to not give any any class advantage especially when using \textit{one vs all}. The kernel used is a Gaussian distribution and is set to be unbounded. To compare the features and their ability to detect the individual seizures, a parallel classification system is needed. Each of the classifiers is trained in a \textit{one vs all} set up. To remove the need for six different confusion matrices, a heatmap was generated for each of the classifiers with the False Positive (FP) and the False Negative (FN) placed at the bottom of the heatmap. The TP value is placed in the predicted and true label of the heatmap the same as a confusion matrix. This makes comparison between each of the individual classifiers easier.
To evaluate the performance of the genetic algorithm on each iteration a NB classifier is used. The NB classifier is faster at classification in comparison to the other machine learning algorithms. When the data is being iterated many times over the same dataset processing speed is critical.
\section{Results and Discussion}\label{Results}
\begin{figure*}[t!]
	\centering
	\includegraphics[width=0.8\textwidth]{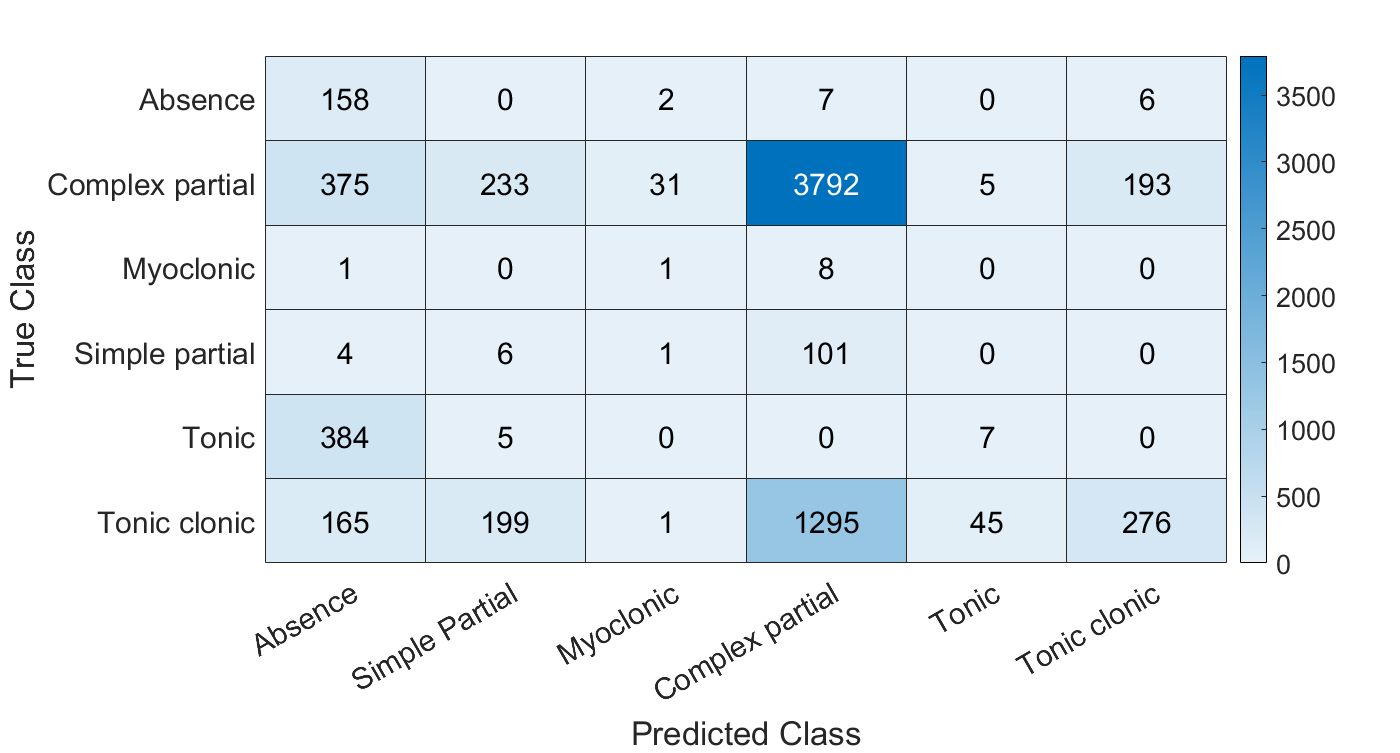}
	\caption{Confusion matrix of the first model.}
	\label{fig:nb_model}
\end{figure*}

The performance metrics used in this state of the art work are shown below
\begin{equation}
	Accuracy= \frac{TP+TN}{TP+TN+FP+FN},
\end{equation}
\begin{equation}
	F1=\frac{TP}{TP+\frac{FP+FN}{2}},
\end{equation}
where $TP,TN,FP$, and $FN$ are the true positive, true negative, false positive, and false negative, respectively.
The first model chosen is the baseline model. This model does not use the genetic algorithm and does not combine the complex partial seizure and simple partial seizure into one label. This model is a single multiclass classifier. The performance of this model is shown in Figure \ref{fig:nb_model}, where there is a high misclassification with the complex partial seizure being classified as simple partial seizures.
It is unexpected that the simple partial and the complex partial are being misdiagnosed as a tonic-clonic seizure. A tonic-clonic seizure can have a focal onset, which can propagate into a bilateral tonic-clonic seizure \cite{ILAERevisedGuide}.

\begin{figure*}[t!]
	\centering
	\includegraphics[width=0.9\textwidth]{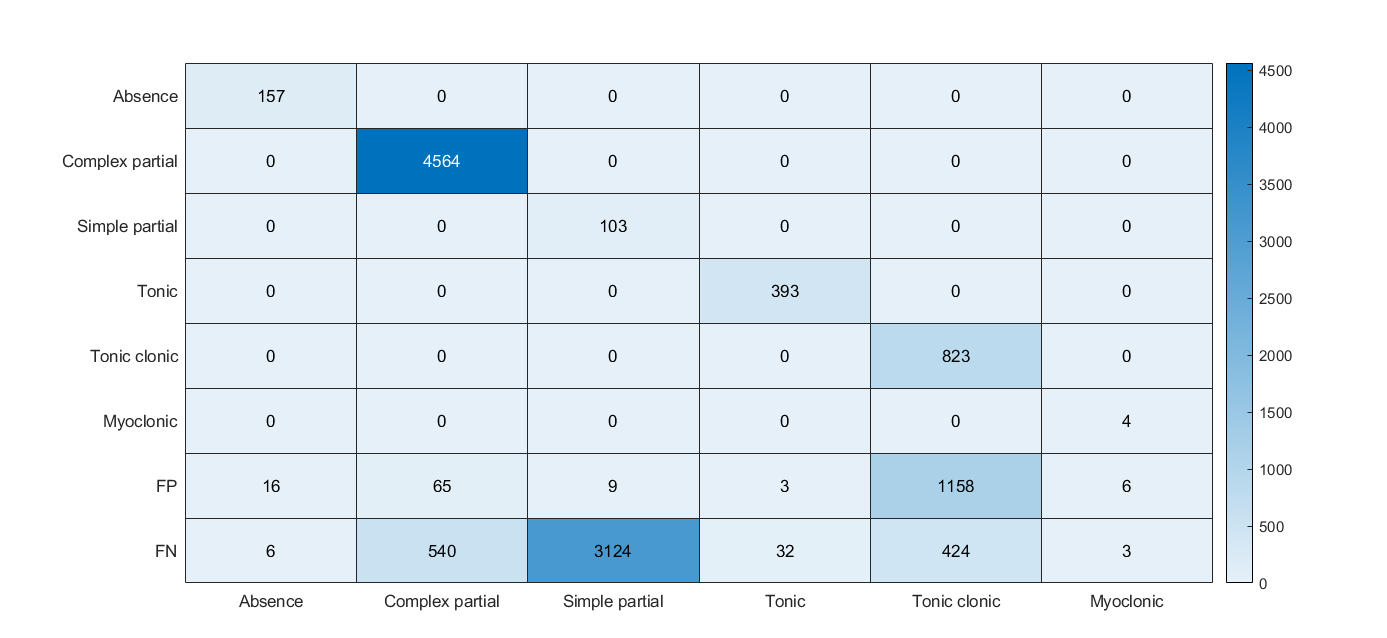}
	\caption{A heatmap describing the performance of the second model.}
	\label{fig:heatmap_adaptions}
\end{figure*}

The second model uses the genetic algorithm to find the features that provide the most optimum F1-score using the labels in the TUH Seizure Corpus. This is a parallel classifier and contains six NB classifiers. Figure \ref{fig:heatmap_adaptions} shows the performance of this model. The tonic seizure had the highest F1-score with $0.96$ followed by the absence seizure with $0.93$. The tonic-clonic and simple partial seizure performed poorly by having higher FPs and FNs than TPs in the system. The heatmap does highlight a large reduction in FPs. There is also an increase in the classification of myoclonic seizures by 400\% compared to Figure \ref{fig:nb_model}.

\begin{figure*}[t!]
	\centering
	\includegraphics[width=0.9\textwidth]{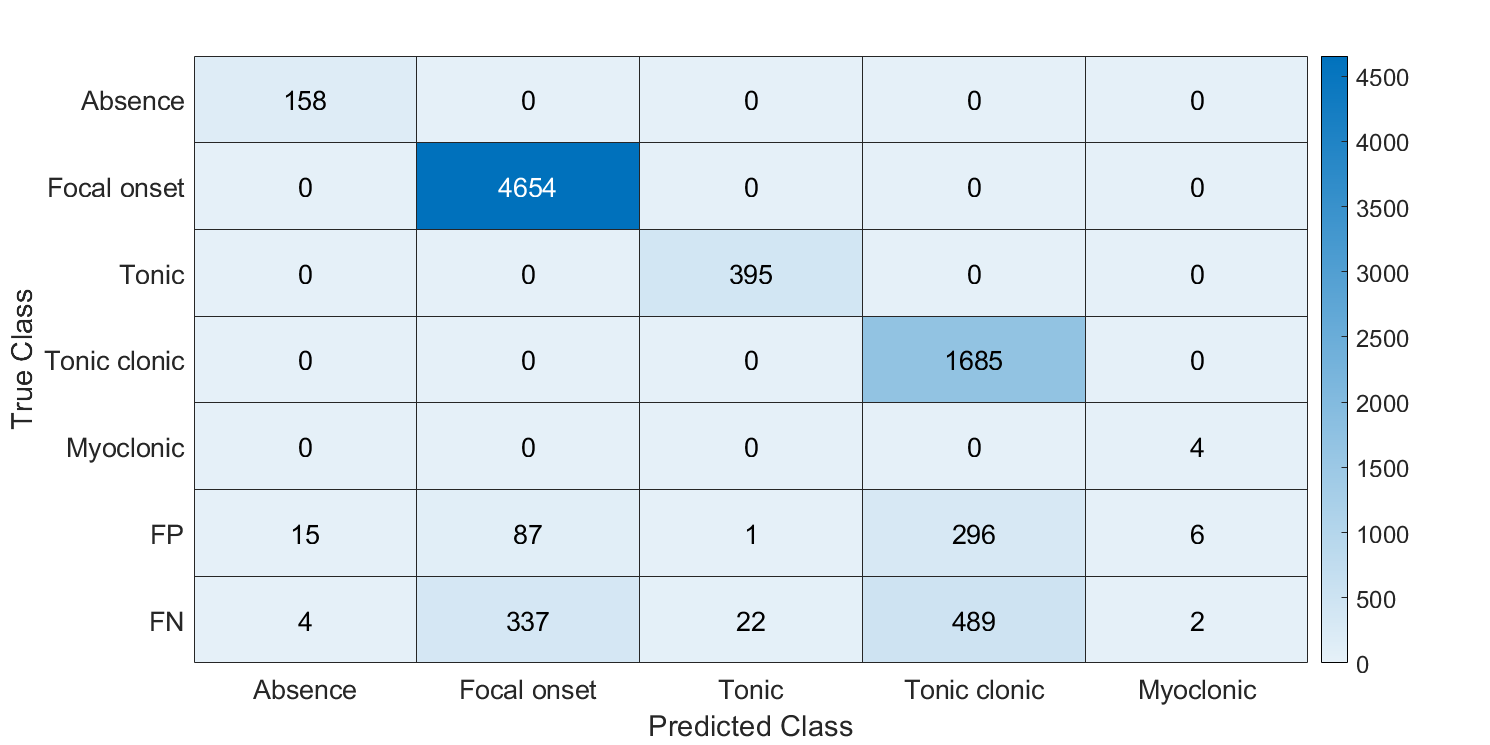}
	\caption{Heatmap describing the proposed PGNBSC classifier with the simple partial and the complex partial seizures combined as focal-onset.}
	\label{fig:nb_heatmap_with_focal}
\end{figure*}

The third model uses the genetic algorithm to find the features that provide the most optimum F1-score with the combined label of both complex partial and simple partial seizures. This is a parallel classifier and contains five NB classifiers. The performance of the third model is shown in Figure \ref{fig:nb_heatmap_with_focal}. Combining the complex partial seizure and the simple partial seizure label has increased the accuracy of the entire system; tonic-clonic seizures had an increase in TPs by 100\% and FPs decreased by 391\% compared to Figure \ref{fig:heatmap_adaptions}. However, FNs of tonic-clonic seizures had increased by 13\%. Also, combining the labels for complex partial and simple partial seizures has had a positive effect on their FNs and FPs, with a combined reduction of 88\%. In Figure \ref{fig:heatmap_adaptions}, the combined TP value for complex partial and simple partial seizures is 4667. However, in Figure \ref{fig:nb_heatmap_with_focal} the combined value is 4654, which is only a decrease of $<1\%$.

In comparison with Kukker et al. \cite{intro_comp_3} the models did not achieve the same accuracy of $96.79$\%. However, the proposed models have improved classification performance by using smaller time windows in the design and the multiclass classification system adopted.


Depending on the chosen classifier, an implementation of a genetic algorithm should be considered for two reasons; firstly to increase the accuracy of classification, and secondly to reduce the size of the feature array.

However, using parallel classifiers for multiclass seizure classification  increases the storage space required for the final implementation in comparison with having a single classifier performing multiclass detection. The processing power is also increased because of the multiple iterations performed. Having said that, this increase in resources requirements is compensated by a higher classification accuracy and an overall more robust classification system. The combination of the focal onset seizures is required to increase the systems accuracy with no real-life effects on the likelihood of the patient.

\section{Conclusion}\label{Conclusion}
The implementation of a genetic algorithm can be crucial when building a machine learning platform to increase classification of seizure types. It requires more training time but further investigations can highlight which features are comparative.
The value of spending increased time at the training stage will ensure that the model increases in accuracy along with combining focal onset seizures to further increase the classification rate.

To further improve the models accuracy, a new set of features that are more tailored to each seizure type can be investigated. More studies can also be carried out to balance the classification performance and the resources requirements factor.

\bibliographystyle{unsrt}
\bibliography{refs}

\end{document}